\documentclass[conference]{IEEEtran}

\usepackage{outlines}
\usepackage{cite}
\usepackage{amsmath,amssymb,amsfonts}
\usepackage{algorithmic}
\usepackage{graphicx}
\usepackage{textcomp}
\usepackage{xcolor}
\def\BibTeX{{\rm B\kern-.05em{\sc i\kern-.025em b}\kern-.08em
    T\kern-.1667em\lower.7ex\hbox{E}\kern-.125emX}}

\usepackage{tikz}
\usepackage{tikz-cd}
\usetikzlibrary{arrows}
\usepackage{tikz-qtree}
\usetikzlibrary{shapes,positioning,chains,fit,calc}
\usetikzlibrary{decorations.markings}

\usepackage[all]{nowidow}

\tikzstyle arrowstyle=[scale=1]
\tikzstyle directed=[postaction={decorate,decoration={markings,
		mark=at position .65 with {\arrow[arrowstyle]{stealth}}}}]
\tikzstyle reverse directed=[postaction={decorate,decoration={markings,
		mark=at position .65 with {\arrowreversed[arrowstyle]{stealth};}}}]

\usepackage{dcolumn}% Align table columns on decimal point
\usepackage{bm}% bold math
\usepackage{hyperref}% add hypertext capabilities
\usepackage{braket} % Added by RS for Bra-Ket notation

\DeclareMathOperator*{\argmin}{arg\,min}

\usepackage{changes}
\definechangesauthor[color=orange]{rs}
\definechangesauthor[color=red]{jl}
\definechangesauthor[color=blue]{is}

\usepackage{braket} % Added by RS for Bra-Ket notation
\usepackage{bm} % added by RS for bold vectors
\newcommand{\vect}[1]{\boldsymbol{\mathbf{#1}}}
\newcommand{\D}{\mathcal{D}}
\newcommand{\HsubM}{\hat{H}_{\!M}}
\newcommand{\HsubC}{\hat{H}_{\!C}}
\newcommand{\hatHsubC}{\hat{H}_{\!C}}

\begin{document}
\bstctlcite{IEEEexample:BSTcontrol} % truncates author list and adds "et al." (see top of qaoa.bib)

\title{Multistart Methods for Quantum Approximate Optimization}% Force line breaks with \\

\author{\IEEEauthorblockN{1\textsuperscript{st} Ruslan Shaydulin}
\IEEEauthorblockA{\textit{School of Computing} \\
\textit{Clemson University}\\
Clemson, South Carolina, USA \\
rshaydu@g.clemson.edu}
\and
\IEEEauthorblockN{2\textsuperscript{nd} Ilya Safro}
\IEEEauthorblockA{\textit{School of Computing} \\
\textit{Clemson University}\\
Clemson, South Carolina, USA \\
isafro@g.clemson.edu}
\and
\IEEEauthorblockN{3\textsuperscript{rd} Jeffrey~Larson}
\IEEEauthorblockA{\textit{Mathematics and Computer Science Division} \\
\textit{Argonne National Laboratory}\\
Lemont, Illinois, USA \\
jmlarson@mcs.anl.gov}
}

\maketitle

\begin{abstract}
        Hybrid quantum-classical algorithms such as the quantum approximate
		optimization algorithm (QAOA) are considered one of the most
		promising approaches for leveraging near-term quantum computers for
		practical applications. Such algorithms are often implemented in a variational form,
		combining classical optimization methods with a
		quantum machine to find parameters to maximize performance.
		The quality of the QAOA solution depends heavily on quality of the parameters produced 
		by the classical optimizer. Moreover, the presence of multiple local optima	
		%\added[id=is]{at this point we already say optimizer/optimization/local optima but not say what is the problem (we actually don't say what is the goal of optimization).}
	    in the space of parameters makes
		it harder for the classical optimizer. In this paper we
		study the use of a multistart optimization approach within
       a QAOA framework to improve the performance of quantum machines on 
		important graph clustering problems. We also demonstrate that reusing
		the optimal parameters from similar problems can improve 
		the performance of classical optimization methods, 
		expanding on similar results
		for MAXCUT.
\end{abstract}

\begin{IEEEkeywords}
quantum approximate optimization, multistart optimization, graph clustering
\end{IEEEkeywords}

% \added[id=rs]{Now available on GitHub at
% \url{https://github.com/rsln-s/Multistart-optimization-for-variational-quantum-algorithms}}

\section{Introduction}

A number of quantum computing devices have recently become
available to researchers~\cite{ballance2016high,barends2014superconducting}.
These Noisy Intermediate Scale Quantum (NISQ) devices currently have 
less than 100 qubits, high error rates, and a restricted set of
available algorithms~\cite{preskill2018quantum}. The famous Shor's
algorithm~\cite{shor1994algorithms} requires executing thousands of
gates~\cite{roetteler2017quantum}, something that is impossible to do
accurately without error correction on quantum machines.
At the same time, there is a growing interest in applying emerging NISQ devices to
practical applications~\cite{shaydulin2018network,shaydulin2018community,ushijima2017graph}.

Multiple near-term algorithms have been introduced in
an attempt to take advantage of NISQ devices. Among the most promising are 
hybrid quantum-classical algorithms, including the Variational Quantum
Eigensolver (VQE)~\cite{peruzzo2014variational} and the Quantum Approximate
Optimization Algorithm (QAOA)~\cite{farhi2014quantum}. These algorithms combine a classical
optimizer with a quantum machine where the quantum evolution is performed
%Quantum evolution is performed
by applying gates to some initial state
(in the case of QAOA, the initial state is an equal superposition of basis states), with 
the goal of preparing the state with desired properties. 
For example, in VQE the goal is to prepare the ground state (i.e., the state
corresponding to the smallest eigenvalue) of a given
system. An
advantage of hybrid algorithms is that the quantum evolution is described by a
shallow-depth circuit, enabling them to be run on NISQ computers
without error correction. The shallow depth of the circuit
is achieved by parameterizing the gates.
An example of a parameterized gate is a rotation around the Z axis (RZ), where
the parameter is the angle of the rotation; the optimal quantum evolution
then can be found by varying the parameters of a shallow set of gates.
In this paper we address 
QAOA, but our approach is applicable to other hybrid
algorithms, including VQE. Although optimal parameters can be found
analytically for some problems, parameters within QAOA
typically are found by using a classical optimizer in a variational
setting. Therefore, in this paper we consider only the variational
implementation of QAOA.

In QAOA, the quantum evolution is performed by applying a series of parameterized
gates, commonly referred to as the ansatz.
These gates are
parameterized by variational parameters, denoted by $\vect{\theta}$. At each step,
a multiqubit trial state $\ket{\psi{(\vect{\theta})}}$ is prepared on the
quantum coprocessor by applying the ansatz. The state is then measured, and the
result is used by the classical optimizer to find new parameters
$\vect{\theta}$, with the goal of finding the ground-state energy
$E_G=\min_{\vect{\theta}}\bra{\psi{(\vect{\theta})}}\HsubC\ket{\psi{(\vect{\theta})}}$,
where $\HsubC$ is the cost Hamiltonian. The ground state encodes the global optimum
of the classical optimization problem. This variational cycle continues until
the classical optimizer converges or a solution of acceptable quality is found.

\begin{figure*}
    \centering
    \hspace{-0.9cm}
    \includegraphics[width=\textwidth]{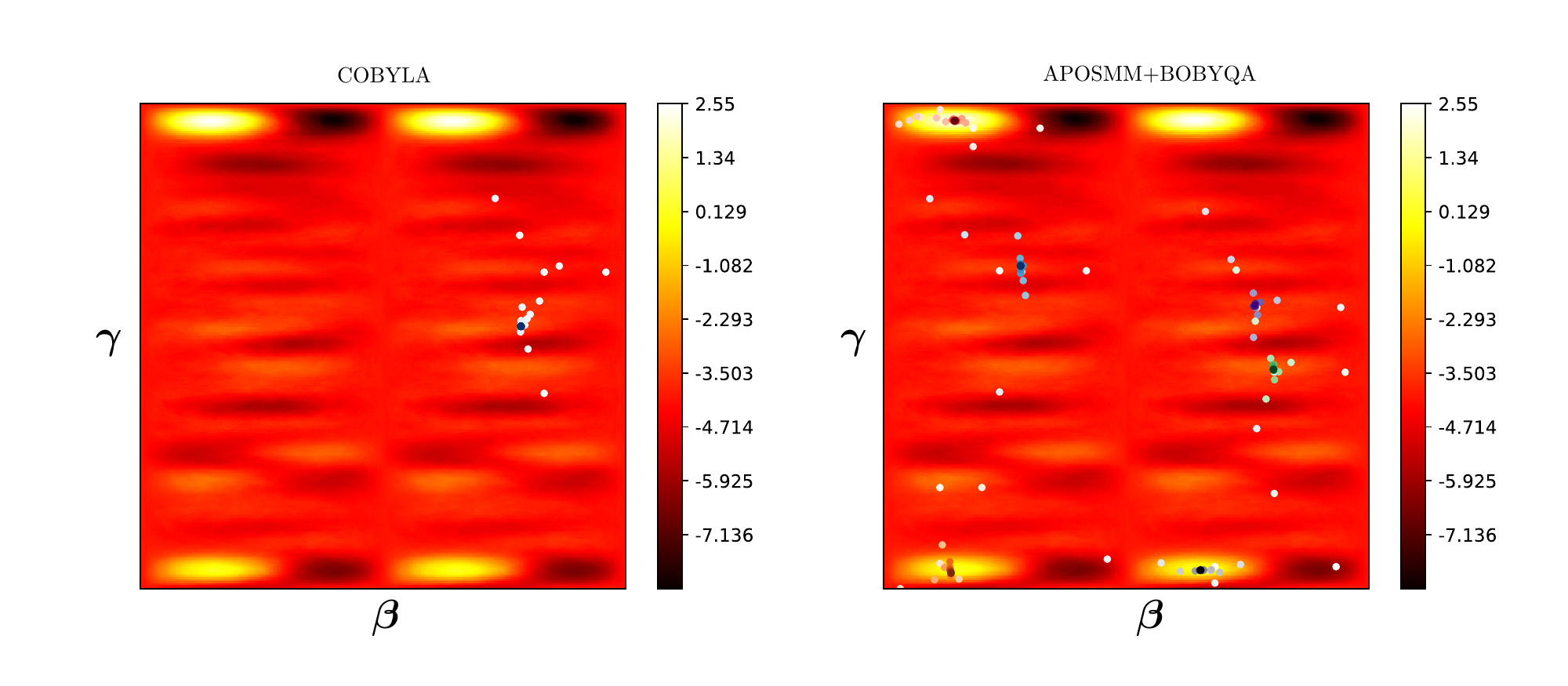}
    \vspace{-0.5cm}
    \caption{Energy landscape of QAOA objective function
    $\bra{\psi{(\beta,\gamma)}}\hatHsubC\ket{\psi{(\beta,\gamma)}}$ for
    modularity maximization community detection on connected caveman
    graph~\cite{watts1999networks, hagberg2008} with 4 cliques of 4 vertices.
    Higher (white) is better. Left: the points evaluated by a single run
    of COBYLA~\cite{powell1994direct,powell1998direct,nlopt}; each point
    corresponds to a pair $(\beta,\gamma)$ that the local optimizer queried. 
    Right: trace of APOSMM~\cite{LarWild14,LW16} coordinating multiple
    COBYLA instances. Both methods were given a budget of 200 function
    evaluations.}
    \label{fig:landscape}
    \vspace{-0.25cm}
\end{figure*}

% The quality of the solution obtained by a variational algorithm depends on the quality of the classical optimization.
% \added[id=is]{Here is what HPEC says about itself: ``We are passionate about performance. Our community is interested in computing hardware, software, systems and applications where performance matters.'' At this point it is still not clear what is the performance problem that we are going to discuss in this paper. This paragraph should be rephrased in terms of performance improvement.}
Such hybrid algorithms are considered the most promising path to demonstrating
quantum advantage, that is, demonstrating superior performance of a quantum
system on some problem when compared with state-of-the-art classical methods.
Demonstrating quantum advantage is a prerequisite for quantum computers to become
a valuable high-performance computing resource. Variational hybrid algorithms,
including VQE and variational implementations of QAOA, require reliable
classical optimization methods to obtain solutions of good quality. Moreover, the
performance of classical optimization methods in terms of the number of
function evaluations directly translates into an improvement in performance of a
variational quantum algorithm. Therefore, it is imperative that efficient and
reliable optimization methods be developed for finding optimal variational
parameters. 
% }{ A reliable classical optimization routine is necessary in order to obtain good
% solutions from a variational algorithm like QAOA. } 
Unfortunately, the parameter space for these problems is 
nonconvex and contains many low-quality,
nondegenerate local optima~\cite{zhou2018quantum}. Figure~\ref{fig:landscape}
shows an example energy landscape of a QAOA objective function 
with two parameters. 
This landscape has many low-quality optima that a local optimizer 
can get stuck in. 
In this paper, we address this challenge by using a multistart local
optimization method. Our results are twofold. First, we explore direct
optimization of QAOA parameters under realistic time constraints and show that
the multistart framework APOSMM~\cite{LarWild14,LW16} is able to find better
parameters than single-start local search methods can (when using the same number of
objective evaluations). Second, we demonstrate that the optimal QAOA parameters found for
a given problem can be reused as an initial point for similar problems,
both improving the quality of the solution and reducing the number of
evaluations required to obtain it.

% libe — good:
% /zfs/safrolab/users/rshaydu/quantum/data/for_jeff/libensemble/modularity_1214_200_2d_get_connected_caveman_graph_left_4_right_4_numparameters_2_noise_False_init_points_20_n_iter_180_seed_6_graph_generator_seed_1_max_active_runs_2_sobol_False.p

% cobyla — bad:

% /zfs/safrolab/users/rshaydu/quantum/data/for_jeff/COBYLA_NLOPT/modularity_1214_200_2d_get_connected_caveman_graph_left_4_right_4_numparameters_2_noise_False_init_points_20_n_iter_180_seed_6_graph_generator_seed_1_max_active_runs_10_sobol_False.p 

% ./plot_heatmap.py /zfs/safrolab/users/rshaydu/quantum/data/heatmaps_ibmqx/1101_get_connected_caveman_graph_left_4_right_4_samples_250_seed_1.p --save --overlay-trace /zfs/safrolab/users/rshaydu/quantum/data/for_jeff/COBYLA_NLOPT/modularity_1214_200_2d_get_connected_caveman_graph_left_4_right_4_numparameters_2_noise_False_init_points_20_n_iter_180_seed_6_graph_generator_seed_1_max_active_runs_10_sobol_False.p --title "COBYLA trace"

QAOA has attracted considerable attention as a candidate algorithm for 
NISQ devices. 
When QAOA was originally introduced in 2014, it was shown to outperform
the state-of-the-art classical solver for the combinatorial
problem of bounded occurrence Max E3LIN2~\cite{farhi2014quantumbounded}. (Thereafter, an
improved classical algorithm was introduced that outperformed QAOA on this problem~\cite{barak2015beating}.) A recent paper~\cite{crooks2018performance}
shows that QAOA (using a circuit with modest depth) can exceed the performance
of Goemans-Williamson~\cite{goemans1995improved} algorithm for MAXCUT. In
addition to these empirical results, 
theoretical results demonstrate that QAOA for MAXCUT
improves on the best-known classical approximation 
algorithms for certain graphs~\cite{osti_1492737, PhysRevA.97.022304}. 
Although there is an active discussion about exactly how many
qubits are required for meaningful quantum
speedups~\cite{Guerreschi2019,shaydulin2018network}, the future of QAOA
looks bright.

\section{Problem Definition}\label{sec:problemdef}

Consider a cost Hamiltonian $\hatHsubC$ encoding the classical
optimization problem (later in this section we present a cost Hamiltonian for
network community detection). Because the
underlying optimization problem we are solving is \emph{maximization}, we construct
% assume that the highest-energy eigenstate of the cost Hamiltonian $\hatHsubC$
the cost Hamiltonian $\hatHsubC$ such that its highest-energy eigenstate 
encodes the solution, as opposed to the ground or lowest-energy state
commonly used in VQE.\footnote{Note that in our case it is just a matter of
convention, since introducing a minus sign changes a maximization problem
into a minimization problem.} 
The goal of the hybrid algorithm is to prepare this eigenstate. In hybrid
quantum-classical algorithms, the evolution is performed by applying
a set of parameterized gates (ansatz). The goal then is to find a set of
parameters that describe the evolution that prepares the desired state. \looseness=-2

In QAOA, the quantum evolution starts in the initial state $\ket{+}^{\otimes
n}$.
Then the evolution is performed by applying two alternating operators
based on the cost Hamiltonian $\hatHsubC$ and mixing Hamiltonian $\HsubM =
\sum_i\hat{\sigma}_i^x$: 
% \added[id=jl]{What are $\hat{sigma}_i^x$? Never appear anywhere else.}
% RS: \hat{sigma}_i^x are operators described Pauli X matrices. It is meaningful
% that we specify them here, since different alterating operators have been 
% explored in the literature. H_M is problem independent, so it doesn't 
% come up again.

\begin{equation}
  \begin{aligned} 
    \ket{\psi{(\vect{\theta})}} & = \ket{\psi{(\vect{\beta},\vect{\gamma})}}\\
    &=  e^{-i\beta_p \HsubM}e^{-i\gamma_p \hatHsubC}\cdots e^{-i\beta_1 \HsubM}e^{-i\gamma_1 \hatHsubC}\ket{+}^{\otimes n}.
  \end{aligned}
\label{eq:ansatz}
\end{equation}

Here $p$ is the number of alternating operators or QAOA ``steps.'' Then the
objective function $f$ (i.e., the energy of $\hatHsubC$ in the state
$\ket{\psi{(\vect{\beta},\vect{\gamma})}}$) is 

\begin{equation}
    f(\vect{\beta},\vect{\gamma}) = -\bra{\psi{(\vect{\beta},\vect{\gamma})}}\hatHsubC\ket{\psi{(\vect{\beta},\vect{\gamma})}}.
    \label{eq:obj}
\end{equation}

Based on the value $f(\vect{\beta},\vect{\gamma})$, the classical 
optimizer chooses the next set of parameters $\vect{\beta},\vect{\gamma}$
with the goal of finding parameters that minimize $f$:

\begin{equation}
\begin{array}{rl}
    \vect{\beta_{*}},\vect{\gamma_{*}} \!\!\!\!\! & = \argmin_{\vect{\beta}, \vect{\gamma}}f(\vect{\beta},\vect{\gamma}) \\
     & = \argmin_{\vect{\beta}, \vect{\gamma}}(-\bra{\psi{(\vect{\beta},\vect{\gamma})}}\hatHsubC\ket{\psi{(\vect{\beta},\vect{\gamma})}}).
\end{array}
\label{eq:argmin_obj_func}
\end{equation}

The objective function $f$ is periodic with respect to $\vect{\beta}$ and
$\vect{\gamma}$, allowing the parameters to be restricted to $\beta_i\in[0,\pi]$, $\gamma_i\in[0,2\pi]$.
Therefore the
optimization domain is compact: $(\vect{\beta},\vect{\gamma})\in
\D=([0,\pi]\times[0,2\pi])^{p}$.

% \added[id=rs]{TODO remove MAXCUT}

% The goal of maximum cut (MAXCUT) problem for undirected graph $G=(V,E)$ is to
% select a set of vertices $S\subset V$ such that the number of edges between $S$
% and $V\\S$ is maximum. The objective function for MAXCUT can be written as
% follows~\cite{brandao2018fixed}:

% \begin{equation}
%     C = \frac{1}{2}\sum_{(ij)\in E}(1-s_is_j)
%     \label{eq:maxcut}
% \end{equation}
% where $A$ is the adjacency matrix of $G$ and $s_i\in\{-1,+1\}$
% indicate whether vertex $i$ is in $S$. MAXCUT is
% NP-hard~\cite{karp1972reducibility} and has numerous practical
% applications~\cite{alidaee19940, neven2008image, deza1994applications}.

% MAXCUT problem can be solved using QAOA by converting spin variables $s_i$ in
% \eqref{eq:maxcut} to Pauli spin operators $\hat{\sigma}^z$:

% \begin{equation}
%     \hatHsubC = \frac{1}{2}\sum_{(ij)\in E}(1-\hat{\sigma}^z_i\hat{\sigma}^z_j)
% \end{equation}

We explore QAOA applied to the modularity maximization problem for the network community
detection.
Also known as graph clustering, 
network community detection aims to group
vertices of the graph so that they
are nontrivially
connected compared with a random graph model. Modularity maximization often (but not necessarily) groups vertices so that there are as many
edges as possible within the groups and as few as possible between the
groups. 
Formally, for an 
undirected graph $G=(V,E)$ with two communities, modularity is defined as
in~\cite{2006PNAS..103.8577N}:
\begin{equation}
C = \frac{1}{4|E|}\sum_{ij}(A_{ij} - \frac{k_ik_j}{2|E|})s_is_j =  \frac{1}{4|E|}\sum_{ij}B_{ij}s_is_j,
\label{eq:mod}
\end{equation}
where $A$ is the adjacency matrix of $G$, $k_i$ is the degree of vertex $i\in V$, and the
variables $s_i\in\{-1,+1\}$ indicate community assignment of vertex $i$. That is, 
$s_i=-1$ denotes vertex $i$ as being assigned to the first community, and $s_j=+1$
denotes that vertex $j$ is assigned to the second community. Modularity
maximization for general graphs is NP-hard~\cite{brandes2006maximizing} and has a variety of
applications in complex
systems~\cite{palla2005uncovering,su2010glay,bardella2016hierarchical,nicolini2017community,negre2018}. \looseness=-2

The modularity maximization problem can be mapped onto QAOA by
promoting variables 
$s_i$ in \eqref{eq:mod} to Pauli spin operators
$\hat{\sigma}^z$~\cite{shaydulin2018network, shaydulin2018community, ushijima2017graph}, resulting in the Hamiltonian
\begin{equation}
    \hatHsubC = \frac{1}{4|E|}\sum_{ij}B_{ij}\hat{\sigma}^z_i\hat{\sigma}^z_j.
\end{equation}

Multiple ansatzes (sets of gates used to produce trial state $\ket{\psi{(\vect{\theta})}}$)
have been explored for QAOA, with the hardware-efficient ansatz~\cite{kandala2017hardware}
(originally proposed for VQE) being one of the most successful~\cite{shaydulin2018network}. A
similar ansatz leveraging nearest-neighbor interactions available on the device
has been shown to achieve a better-than-random-guess approximation ratio
for the MAXCUT problem on 3-regular graphs~\cite{farhi2017quantum}. In this work we
do not consider these ansatzes, however, because at the time of writing there is no evidence that QAOA
with such ansatzes can beat the best classical algorithms. Instead, we focus on the alternating operator ansatz in \eqref{eq:ansatz}.

\section{Related Work}

While the most commonly used strategy for identifying optimal QAOA parameters is using
a classical optimizer in a variational loop, QAOA is not necessarily 
a variational algorithm. For example,
Parekh et al.~show that for 
one-step ($p=1$) QAOA for MAXCUT on $k$-regular triangle-free graphs,
parameters can be derived analytically~\cite{osti_1492737}.
Wang et al. show
a similar
result for one-dimensional antiferromagnetic 
rings~\cite{PhysRevA.97.022304}. More generally, 
Farhi et al.~\cite{farhi2014quantum} proposed discretizing parameters 
into a grid. For $N$-qubit QAOA, however, this approach requires $N^{O(p)}$ 
evaluations, making it impractical even for small $p$. Finding good 
QAOA parameters remains a challenging problem, which motivates this work.

\subsection{Parameter optimization in hybrid algorithms}

% This paragraph can be extended if needed

Despite the recent advances in gradient-based
methods~\cite{guerreschi2017practical, bergholm2018pennylane,
romero2018strategies, zhou2018quantum, crooks2018performance, harrow2019low}, 
gradient-free black-box methods remain the most common approach
for optimizing parameters in hybrid quantum-classical algorithms. A variety of methods have
been used, 
including the Nelder-Mead method~\cite{nelder1965simplex}
(for both QAOA parameter optimization~\cite{Guerreschi2019, guerreschi2017practical}
and training quantum Boltzmann machines~\cite{verdon2017quantum}), Bayesian
methods~\cite{otterbach2017unsupervised}, Powell's method~\cite{wecker2016training}, and
an interior-point minimization method~\cite{yang2017optimizing}. 
Researchers resort to
derivative-free methods because 
analytic gradients for quantum
circuits may not be available and approximating gradients can be computationally
expensive~\cite{crooks2018performance}. (In some cases, algorithmic
differentiation techniques may provide gradient
information~\cite{bergholm2018pennylane}.) Since 
gradient-based methods can be sensitive to 
noise~\cite{zhu2018training}, they may be less suitable for 
noisy intermediate-scale quantum hardware. 

A number of recent advances in finding good parameters 
have been made in the recent years, potentially making their 
optimization simpler. For QAOA, multiple papers have shown 
connections between adiabatic 
schedule and QAOA 
parameters~\cite{brandao2018fixed, crooks2018performance, zhou2018quantum}.  \looseness=-2

Zhou et al.~\cite{zhou2018quantum} show that even at small 
depth $p$ 
the schedule defined by optimal QAOA parameters 
is reminiscent of adiabatic quantum annealing, where $\hatHsubC$
is gradually turned on while $\HsubM$ is gradually turned off 
(see Sec.~\ref{sec:problemdef}). Similar results were found by 
Crooks~\cite{crooks2018performance}.
Additionally, Zhou et al.~\cite{zhou2018quantum} demonstrate
that the optimal values $\vect{\beta_{*}},\vect{\gamma_{*}}$
have small variation between similar problem instances, a finding
that we confirm in this work for a different graph problem 
(see Sec.~\ref{sec:reusing}). 
Zhou et al.~use these insights to introduce a novel parameterization of QAOA and
a heuristic optimization scheme based on it.

Brandao et al.~\cite{brandao2018fixed} show that for
MAXCUT on 3-regular graphs, the objective function value
is concentrated; that is, typical instances have nearly the same value
of objective function. They make a case that the same holds
for any combinatorial search problem where the number
of clauses with a given variable is bounded (e.g., MAXCUT
on a bounded-degree graph). They propose reusing optimal
parameters between problems that come from the same distribution
and refining them using a local optimization heuristic.
In this work, we successfully apply this strategy to modularity clustering,
a problem where the number of clauses in which a variable can appear 
grows with $n$ (see Sec.~\ref{sec:reusing}).

Periodicity of the objective function with respect to QAOA parameters, visible
on heatmaps in 
Fig.~\ref{fig:landscape}, has been demonstrated for 
MAXCUT~\cite{PhysRevA.97.022304, zhou2018quantum}. 
The periodicity was also observed for quasi-maximum-likelihood
decoding of classical channel codes~\cite{matsumine2019channel}.
This can potentially allow for further restriction of the domain,
eliminating some of the local optima and making the optimization
problem easier.
However, the theoretical results so far are problem specific.
Therefore, we restrict our optimization domain 
to $\beta_i\in [0,\pi]$, $\gamma_i\in [0,2\pi]$, following~\cite{farhi2014quantum}.
Note that this differs from the approach in~\cite{crooks2018performance},
where the values of $\vect{\beta}$ and $\vect{\gamma}$ were not constrained.
A recent result shows that exploiting the periodicity of 
variational parameters of certain ansatzes for QAOA and other 
variational algorithms can improve optimization
performance~\cite{nakanishi2019sequential}. % omg do I really not want to talk about that paper

\subsection{Derivative-free optimization methods}

Selecting $\vect{\beta}$ and $\vect{\gamma}$ values that maximize the objective function in
\eqref{eq:argmin_obj_func} is a central optimization problem in
variational algorithms.
Since the gradient of the objective function with respect to $\vect{\beta}$ and $\vect{\gamma}$
is unavailable
on real quantum computers, 
researchers usually resort to so-called derivative-free optimization (DFO) methods:
those that work only with observations of the objective function. 
Classical derivative-free direct-search methods are commonly applied to such
problems: for example, Nelder-Mead is the default method for VQE problems in
Grove \cite{Grove}. Yet McClean et al.~\cite{mcclean2016theory} shows that modern DFO methods
achieve considerable benefits in terms of the number of function evaluations
required. The BOBYQA method \cite{Powell2009a} is one such method for
bound-constrained derivative-free optimization that builds quadratic models of
the objective and optimizes them over a trust region in order to produce
candidate points.

In the numerical optimization community, one commonly starts local
optimization methods from different initial conditions in an attempt to identify
better optima.  While such an approach
%starting a local optimization run from each point in a sample
may be easy to implement, it may result in unnecessary
objective function evaluations. Assuming there are a finite number of local
optima, the ideal approach would identify each using only a single local run.

The multilevel single linkage method (MLSL) \cite{RinnooyKan1987,RinnooyKan1987a},
uniformly samples points over the domain $\D$ and starts runs from those points that do not have a better point
within a ball of some radius. They show favorable results for a specific approach for
updating the radius as the number of sampled points increases, although such results are only asymptotic. 
MLSL was generalized by APOSMM~\cite{LarWild14,LW16} to consider all
points generated by an ensemble of local optimization runs, and not just those
sampled from the domain. 

\section{Difficulty of Optimizing QAOA Parameters}\label{sec:difficulty}

\begin{figure*}
    \centering
    \hspace{-0.3cm}
    \includegraphics[width=\linewidth]{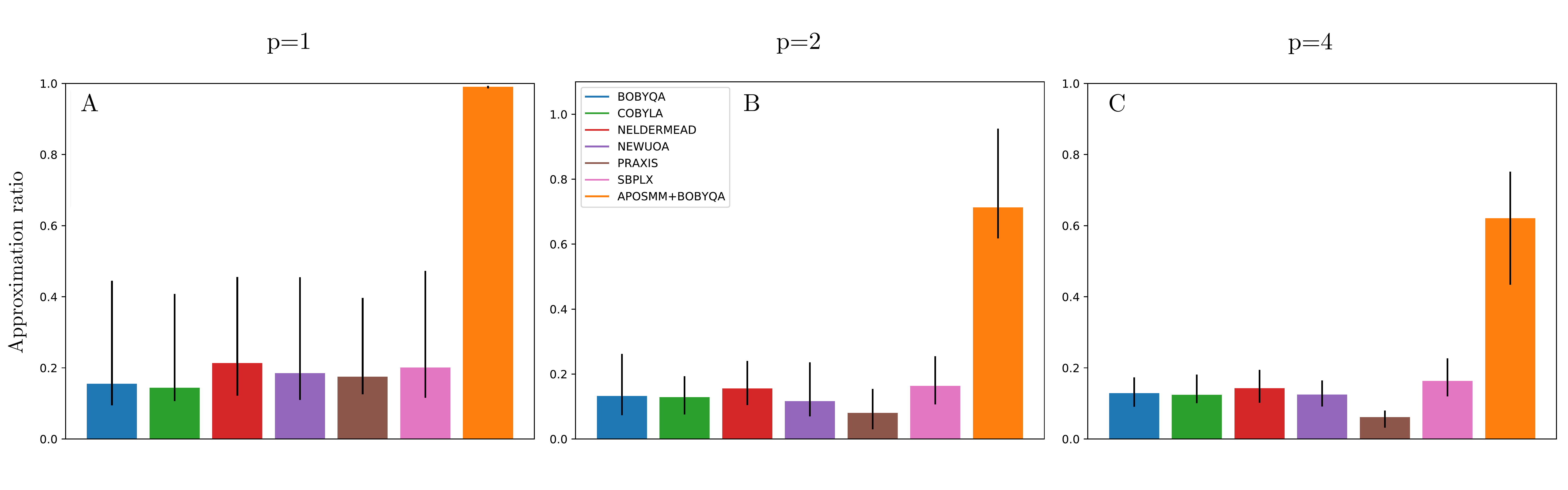}
    \vspace{-0.5cm}
    \caption{Ratio between the value of the objective function found by an optimization
    method and the best-found value. All local methods are run with \emph{no}
    restart and zero tolerances. Heights of bars represent median over
    $\mbox{(10 seeds per problem)}\times\mbox{(6 problems)}=60$ runs. Error
    bars represent quartiles (25th and 75th percentiles). When compared with
    local methods without restarting, APOSMM finds solutions with much higher
    objective function values. This is due to local methods converging before
    exhausting the budget on number of function evaluations. $p$ is number of
    QAOA steps ($p=1$ corresponds to 2-dimensional domain $\D$ (A), $p=2$
    corresponds to $dim(\D)=4$ (B), and $p=4$ corresponds to $dim(\D)=8$ (C).) Note that
    the approximation ratio $1.0$ corresponds to the maximum value observed for a given
    problem and given value of $p$. The maximum absolute values of the objective
    function vary between the different numbers of QAOA steps.}
    \label{fig:norestart_bar}
    \vspace{-0.25cm}
\end{figure*}

\begin{figure*}
    \centering
    \hspace{-0.2cm}
    \includegraphics[width=\textwidth]{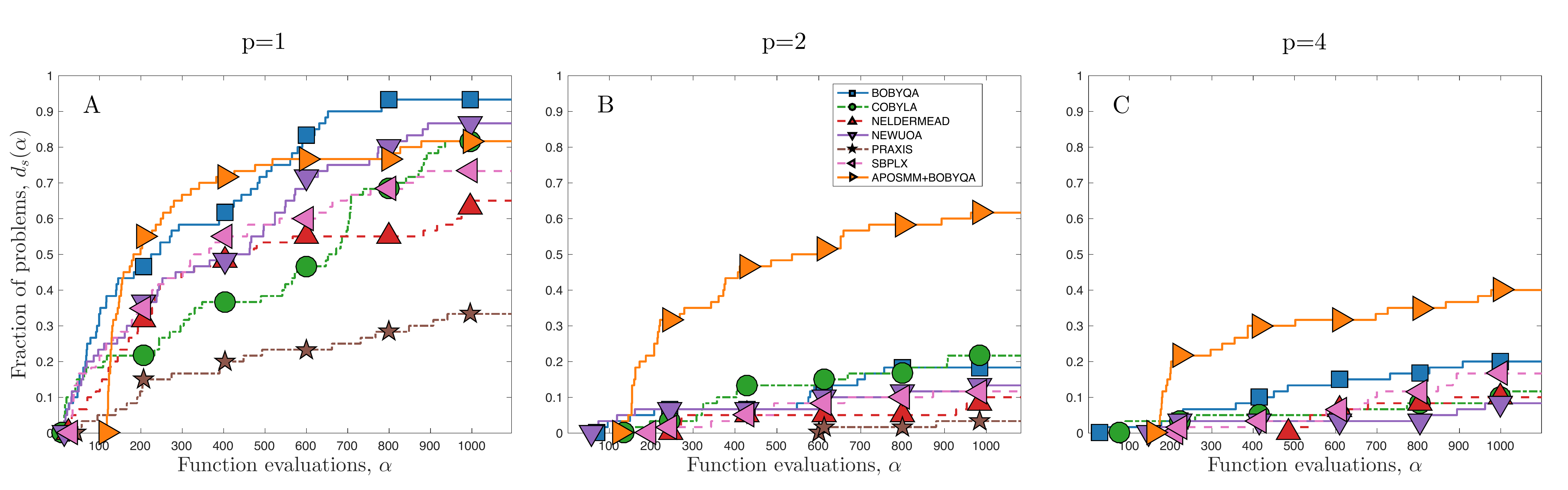}
    \vspace{-0.5cm}
    \caption{Data profiles for seven optimization methods on the $p=1$ (A),
    $p=2$ (B), and $p=4$ (C) benchmark problems with $\tau=0.01$. For the $p=1$ (i.e., two-dimensional)
    problems, most methods are competitive; but as the number of parameters
    (i.e., circuit depth) increases, all methods have difficulty in identifying
    high-quality solutions on a large fraction of the test problems. Yet, we
    see that APOSMM+BOBYQA performs noticeably better.
    \label{fig:restart_profiles}}
    \vspace{-0.25cm}
\end{figure*}

In this section we present the results from using DFO methods to find optimal
QAOA parameters. We use the high-performance simulator Qiskit Aer~\cite{Qiskit} to perform noiseless simulations of QAOA circuits. We measure the quality of the solution found by six
derivative-free local optimization methods as implemented in the NLopt
nonlinear-optimization package~\cite{nlopt}:
BOBYQA~\cite{powell2009bobyqa}, 
COBYLA~\cite{powell1994direct,powell1998direct},
NEWUOA~\cite{powell2006newuoa}, 
Nelder-Mead~\cite{nelder1965simplex},
PRAXIS~\cite{brent2013algorithms} and SBPLX~\cite{rowan1991functional}. We
compare their performance to the implementation of APOSMM
from the libEnsemble library~\cite{libEnsemble_0.5.0}.
APOSMM coordinates multiple local optimization runs in an attempt to identify better local optima.
In this work, we use BOBYQA as the local optimization method within APOSMM (we
denote this method as APOSMM+BOBYQA in figures).
The performance of all methods is evaluated using two-way modularity maximization community
detection problem on six synthetic graphs with community structure: three
instances of connected caveman graph~\cite{watts1999networks} and three
instances of random partition graph~\cite{fortunato2010community}. All graphs
have between 10 and 12 vertices and were generated with
NetworkX~\cite{hagberg2008}. The code used to perform the experiments is available~\cite{code}.

We performed two sets of experiments. First, we set the tolerances of local
solvers to zero and allow them to run until convergence. The quality of the
obtained solutions was then compared with the solutions found by APOSMM using
the same number of evaluations. We observe that APOSMM finds solutions with
a much higher value of the objective function (see Fig. \ref{fig:norestart_bar}).
Since APOSMM is allowed another local optimization run after one has converged,
a local
method may not take full advantage of the function evaluations budget. To
allow for a more equal comparison, we performed a second set of experiments. In the second set, we
set the tolerances of local solvers to be equal to the tolerance of BOBYQA
within APOSMM and if the method convergence before exhausting function
evaluations, it is restarted at a different random point. This restart scheme
is essentially a naive version of MLSL. These results are also compared with
APOSMM (see Fig.~\ref{fig:restart_profiles}).

For both sets of the experiments, we limit the number of evaluations to 1,000. We
choose this number as the realistic number of evaluations based on the
estimates in~\cite{Guerreschi2019}. We use the same realistic if
aggressive assumption of 1 millisecond per single run. Estimating objective function
in Eq.~\ref{eq:obj} requires thousands to tens of thousands measurements in practice~\cite{Guerreschi2019, kandala2017hardware, otterbach2017unsupervised}; we use an optimistic
assumption of 1,000 measurements per run for obtaining the statistics to
estimate the objective function. %\added[id=jl]{Please clarify this previous sentence.}
This gives an estimate on the time cost of
performing the optimization equal to $\mbox{(time per single
run)}\times\mbox{(1,000 measurements per run)}\times\mbox{(1,000
evaluations)}\approx 16\mbox{ min}$. Note that this runtime is still orders of
magnitude greater than the runtime of classical state-of-the-art MAXSAT
solvers applied to the same problem~\cite{Guerreschi2019}. Additionally, as
the hardware is rapidly evolving, it is not possible to project these estimates
into the future with certainty. However, it provides a useful estimate on the
reasonable number of calls to the quantum device in a QAOA run. 

Results show that a single run of a local optimization method cannot identify
parameters $(\vect{\beta},\vect{\gamma})$ corresponding to a high-quality solution of
underlying problem (i.e., a high value of objective function).
Fig.~\ref{fig:norestart_bar} shows that APOSMM is capable of finding parameters
corresponding to values of objective function much larger than just the local
solvers. This is partly due to local solvers converging before exhausting the
limit on number of function evaluations (1,000).

If we set tolerances for local methods to the same values as in APOSMM (the
tolerances on change in the function value to $10^{-3}$ and on the change in
optimization parameters to $10^{-2}$) and restart local methods after
convergence, we observe that APOSMM is still solving more problems within the same
budget of function evaluations. This is measured in the data profiles in 
Fig.~\ref{fig:restart_profiles}; these data profiles track the
fraction of problems solved to some level $\tau$ after a given number of
function evaluations. Explicitly, if $t_{p,s}$ is the number of function
evaluations required for each optimization method $s$ to solve problem $p$ in
the set of problems $P$, then the data profile is
\[
d_s(\alpha) = \frac{\left| \left\{ p : t_{p,s} \le \alpha 
\right\}  \right|}{\left| P \right|}.
\]
where $\alpha$ is the number of function evaluations.  
Data profiles require some definition of solving a problem to a level $\tau$. For
these problems, an optimization method $s$ is determined to
have solved problem $p$ to a level $\tau$ after $j$ evaluations if  
\begin{equation}\label{eq:convergence_f}
  f(x^0) - f(x^j) \ge (1-\tau)(f(x^0) - \tilde{f}_p),
\end{equation}
% \added[id=jl]{Shoudl the sign of this be changed as we are maximizing?}
% RS: No, since the problem in Eq. 3 is minimization and that is the f that optimization 
% methods are working with. Same below.
where $x^0$ is the problem's starting point, $x^j$ is the
$j$th point evaluated by the method, and
$\tilde{f}_p$ is the best-found
function value by any optimization method on problem $p$. 
For example, if $\tau = 0.01$, the convergence test in~\eqref{eq:convergence_f}
determines a method to solve problem $p$ when a point is evaluated
with 99\% of the possible decrease %\added[id=jl]{increased?}
on the problem (among the implementations
being compared).

Figures \ref{fig:norestart_bar} and \ref{fig:restart_profiles} demonstrate that
finding optimal parameters becomes increasingly harder as the
dimension of the domain $\D$ (i.e., the number of QAOA steps $p$) increases.
For $p=1$, BOBYQA and APOSMM solve most of the problems
(Fig.~\ref{fig:restart_profiles}A) within 1,000 function evaluations, for $p=2$
and $p=4$ the best-performing method (APOSMM) solves only 60\% and 40\% of the
problems, respectively. These results indicate that even for small number of QAOA steps
($p=4$) direct optimization of variational parameters is hard under
realistic time constraints.

\section{Reusing Optimal QAOA Parameters}\label{sec:reusing}

\begin{figure}
    \centering
    \hspace{-0.3cm}
    \includegraphics[width=\linewidth]{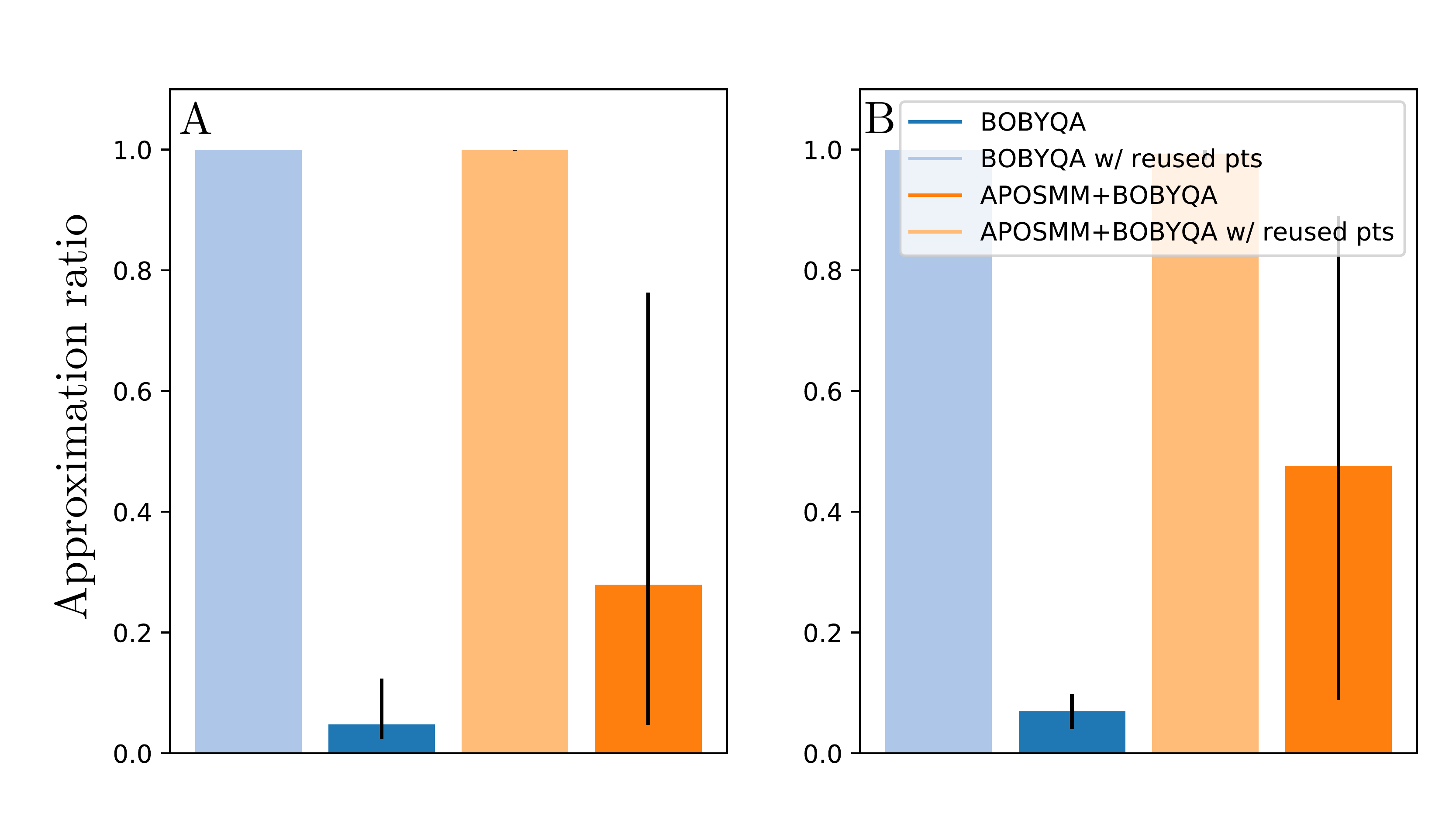}
    \vspace{-0.3cm}
    \caption{Ratio between the value of the objective function found by an optimization
    method and the best-found value. Left (A): we compare the best-performing
    local method and APOSMM with optimal points from similar problems (``w/
    reused pts'') and with random initial points. Heights of bars represent
    median over $\mbox{(10 seeds per problem)}\times\mbox{(6
    problems)}\times\mbox{(5 different random edges removed)}=300$ runs. 
    Right (B): for each problem we remove only one ``worst-case'' edge. Error
    bars represent quartiles (25th and 75th percentiles). Reusing precomputed
    optimal points allows optimization methods to find better solutions
    (corresponding to higher objective values) within the same
    budget of function evaluations.}
    \label{fig:remove_edge_bar}
    \vspace{-0.25cm}
\end{figure}

Sec.~\ref{sec:difficulty} presents results demonstrating the complexity of
finding good QAOA parameters under realistic time constraints. Recently a
number of researchers proposed amortizing the cost of finding good QAOA
parameters for MAXCUT by reusing optimal parameters found for a given problem
on similar problems~\cite{crooks2018performance, zhou2018quantum,
brandao2018fixed}. We confirm and extend these findings by reusing optimal QAOA
parameters found by exhaustive search. Optimal parameters for QAOA for
modularity maximization on a given graph are used as an initial guess for the local
solver on a similar graph constructed by removing an
edge from the original graph. This simulates a realistic scenario of solving
community detection on a dynamical graph, for example, a social network where new
friend connections are dynamically added and removed. 

We estimate true optimal parameters by setting the tolerance on the change in the function
value to $10^{-3}$ and the tolerances on the change in the parameters to
$10^{-2}$ and restarting BOBYQA after each convergence until 100,000 function
evaluations have been used. We observe
that this exhaustive approach identifies multiple high-quality local optima. We
then use these high-quality QAOA parameters as initial guesses for local
methods and APOSMM. After a local method converges, it is restarted from the
next-best local optima.

Our contribution extends previous work in two ways. First, we consider a
different optimization problem, namely, modularity community detection. Second,
in addition to random similar problems, we consider ``worst-case'' small changes.
To simulate a ``worst-case'' scenario, we remove an edge from the graph that has
the greatest impact on its spectrum. Concretely, we compute the spectrum of the
graph Laplacian before and after removing an edge. The change in the spectrum is
measured by computing the Euclidean distance between the eigenvectors of the
graph Laplacians. The graph spectrum has deep connections to many optimization
problems on graphs, including graph partitioning and community detection~\cite{chung1997spectral,nadakuditi2012graph}.
%\added[id=rs]{citations needed (@Ilya?)}.

Figure \ref{fig:remove_edge_bar} presents the results. We observe that using
optimal parameters from similar problems allows optimization methods to find
high-quality solutions under realistic time constraints.
Thus, 
we are hopeful
that the high cost of finding good QAOA parameters can be amortized by reusing the
parameters from similar problems.

\section{Discussion}

This paper present results on finding optimal QAOA parameters to improve the performance of quantum 
optimization solvers. We show
that multistart methods such as APOSMM can utilize a fixed number of function
evaluations more efficiently by interleaving multiple local optimization runs
and considering all $(\vect{\beta},\vect{\gamma})$ parameters generated by them.
We observe that as the number of
QAOA steps and the dimension of the corresponding optimization domain $\D$ is
increased, the optimization problem becomes increasingly hard. These results highlight
the need to develop more efficient approaches to finding optimal parameters 
to accelerate and improve the performance of QAOA---a challenge because, in order 
to compete with state-of-the-art classical
solvers on problems with fewer than 200 variables, QAOA has to run in no more
than a minute~\cite{Guerreschi2019, shaydulin2018network}. An additional
challenge is presented by the high levels of noise on near-term hardware. 

We show that the obstacles can be partially addressed by reusing optimal
parameters found for a
similar problem. We observe that parameters can be reused both for similar
problems with a random change introduced
and in ``worst-case'' scenarios, where the change in
the underlying problem has the greatest impact on its structure. For example,
reusing optimal parameters found for $p=1$ using BOBYQA or APOSMM for a dynamic
graph over 1,000 changes and allowing local methods a realistic 10--30 iterations
in order to refine reused optimal points at each iteration, would bring amortized cost
down from $\mbox{(1 ms per run)}\times\mbox{(1,000 measurements per
run)}\times\mbox{(1,000 evaluations)}\approx	16\mbox{ min}$ minutes to a more
competitive $\approx 1$ second. Reusing parameters and employing heuristic
techniques such as FOURIER proposed in~\cite{zhou2018quantum} could bring
down amortized costs of QAOA run even further. We believe this could make quantum
optimization solvers a valuable extreme-computing resource.

The limited connectivity between qubits in many hardware implementations 
presents an additional challenge. For example, superconducting qubit
technology, developed by, among others, IBM, Rigetti, and Google,
provides only nearest-neighbor connectivity with qubits arranged on 
a two-dimensional lattice.
The modularity maximization graph clustering problem discussed in this paper
requires all-to-all connectivity.
The connectivity limitation can be addressed by a SWAP 
network~\cite{anschuetz2019variational, Babbush2018, crooks2018performance}
with only $O(N)$ overhead (where $N$ is the number of qubits).
Additionally, ion-trap architectures (the most famous implementation developed
by IonQ) do not have the same connectivity limitations, because they allow
the application of gates between any pair of qubits.

All these factors strengthen the potential of QAOA. As hardware
continues to improve and more advanced techniques for parameter optimization
are developed, QAOA has the potential to outperform classical state-of-the-art solvers.

\section*{Acknowledgments}
The authors would like to
acknowledge Yuri Alexeev for insightful discussions. 
This material was based upon work supported by the U.S. Department of Energy,
Office of Science, under contract DE-AC02-06CH11357. 
Clemson University is acknowledged for generous allotment of compute time on Palmetto cluster.

\bibliographystyle{IEEEtran}
\bibliography{qaoa}

\begin{thebibliography}{10}
\providecommand{\url}[1]{#1}
\csname url@samestyle\endcsname
\providecommand{\newblock}{\relax}
\providecommand{\bibinfo}[2]{#2}
\providecommand{\BIBentrySTDinterwordspacing}{\spaceskip=0pt\relax}
\providecommand{\BIBentryALTinterwordstretchfactor}{4}
\providecommand{\BIBentryALTinterwordspacing}{\spaceskip=\fontdimen2\font plus
\BIBentryALTinterwordstretchfactor\fontdimen3\font minus
  \fontdimen4\font\relax}
\providecommand{\BIBforeignlanguage}[2]{{%
\expandafter\ifx\csname l@#1\endcsname\relax
\typeout{** WARNING: IEEEtran.bst: No hyphenation pattern has been}%
\typeout{** loaded for the language `#1'. Using the pattern for}%
\typeout{** the default language instead.}%
\else
\language=\csname l@#1\endcsname
\fi
#2}}
\providecommand{\BIBdecl}{\relax}
\BIBdecl

\bibitem{ballance2016high}
C.~Ballance, T.~Harty, N.~Linke, M.~Sepiol, and D.~Lucas, ``High-fidelity
  quantum logic gates using trapped-ion hyperfine qubits,'' \emph{Physical
  Review Letters}, vol. 117, no.~6, p. 060504, 2016.

\bibitem{barends2014superconducting}
R.~Barends, J.~Kelly, A.~Megrant, A.~Veitia, D.~Sank \emph{et~al.},
  ``Superconducting quantum circuits at the surface code threshold for fault
  tolerance,'' \emph{Nature}, vol. 508, no. 7497, p. 500, 2014.

\bibitem{preskill2018quantum}
J.~Preskill, ``Quantum computing in the {NISQ} era and beyond,''
  \emph{arXiv:1801.00862}, 2018.

\bibitem{shor1994algorithms}
P.~W. Shor, ``Algorithms for quantum computation: Discrete logarithms and
  factoring,'' in \emph{Proceedings 35th Annual Symposium on Foundations of
  Computer Science}.\hskip 1em plus 0.5em minus 0.4em\relax IEEE, 1994, pp.
  124--134.

\bibitem{roetteler2017quantum}
M.~Roetteler, M.~Naehrig, K.~M. Svore, and K.~Lauter, ``Quantum resource
  estimates for computing elliptic curve discrete logarithms,'' in
  \emph{International Conference on the Theory and Application of Cryptology
  and Information Security}.\hskip 1em plus 0.5em minus 0.4em\relax Springer,
  2017, pp. 241--270.

\bibitem{shaydulin2018network}
R.~Shaydulin, H.~Ushijima-Mwesigwa, I.~Safro, S.~Mniszewski, and Y.~Alexeev,
  ``Network community detection on small quantum computers,'' \emph{{A}dvanced
  {Q}uantum {T}echnologies (to appear), ar{X}iv:1810.12484}, 2019.

\bibitem{shaydulin2018community}
------, ``Community detection across emerging quantum architectures,''
  \emph{Proceedings of the 3rd International Workshop on Post Moore's Era
  Supercomputing}, 2018.

\bibitem{ushijima2017graph}
H.~Ushijima-Mwesigwa, C.~F. Negre, and S.~M. Mniszewski, ``Graph partitioning
  using quantum annealing on the {D-Wave} system,'' in \emph{Proceedings of the
  Second International Workshop on Post Moore's Era Supercomputing}.\hskip 1em
  plus 0.5em minus 0.4em\relax ACM, 2017, pp. 22--29.

\bibitem{peruzzo2014variational}
A.~Peruzzo, J.~McClean, P.~Shadbolt, M.-H. Yung, X.-Q. Zhou \emph{et~al.}, ``A
  variational eigenvalue solver on a photonic quantum processor,'' \emph{Nature
  Communications}, vol.~5, p. 4213, 2014.

\bibitem{farhi2014quantum}
E.~Farhi, J.~Goldstone, and S.~Gutmann, ``A quantum approximate optimization
  algorithm,'' \emph{arXiv:1411.4028}, 2014.

\bibitem{zhou2018quantum}
L.~Zhou, S.-T. Wang, S.~Choi, H.~Pichler, and M.~D. Lukin, ``Quantum
  approximate optimization algorithm: Performance, mechanism, and
  implementation on near-term devices,'' \emph{arXiv:1812.01041}, 2018.

\bibitem{LarWild14}
J.~Larson and S.~M. Wild, ``A batch, derivative-free algorithm for finding
  multiple local minima,'' \emph{Optimization and Engineering}, vol.~17, no.~1,
  pp. 205--228, 2016.

\bibitem{LW16}
------, ``Asynchronously parallel optimization solver for finding multiple
  minima,'' \emph{Mathematical Programming Computation}, vol.~10, no.~3, pp.
  303--332, 2018.

\bibitem{watts1999networks}
D.~J. Watts, ``Networks, dynamics, and the small-world phenomenon,''
  \emph{American Journal of Sociology}, vol. 105, no.~2, pp. 493--527, 1999.

\bibitem{hagberg2008}
A.~A. Hagberg, D.~A. Schult, and P.~J. Swart, ``Exploring network structure,
  dynamics, and function using {NetworkX},'' in \emph{Proceedings of the 7th
  Python in Science Conference (SciPy 2008)}, G.~Varoquaux, T.~Vaught, and
  J.~Millman, Eds., Pasadena, CA USA, 2008, pp. 11--15.

\bibitem{powell1994direct}
M.~J. Powell, ``A direct search optimization method that models the objective
  and constraint functions by linear interpolation,'' in \emph{Advances in
  Optimization and Numerical Analysis}.\hskip 1em plus 0.5em minus 0.4em\relax
  Springer, 1994, pp. 51--67.

\bibitem{powell1998direct}
M.~Powell, ``Direct search algorithms for optimization calculations,''
  \emph{Acta Numerica}, vol.~7, pp. 287--336, 1998.

\bibitem{nlopt}
\BIBentryALTinterwordspacing
S.~G. Johnson, ``The {NLopt} nonlinear-optimization package,'' 2019. [Online].
  Available: \url{http://github.com/stevengj/nlopt}
\BIBentrySTDinterwordspacing

\bibitem{farhi2014quantumbounded}
E.~Farhi, J.~Goldstone, and S.~Gutmann, ``A quantum approximate optimization
  algorithm applied to a bounded occurrence constraint problem,''
  \emph{arXiv:1412.6062}, 2014.

\bibitem{barak2015beating}
B.~Barak, A.~Moitra, R.~O'Donnell, P.~Raghavendra, O.~Regev \emph{et~al.},
  ``Beating the random assignment on constraint satisfaction problems of
  bounded degree,'' \emph{arXiv:1505.03424}, 2015.

\bibitem{crooks2018performance}
G.~E. Crooks, ``Performance of the quantum approximate optimization algorithm
  on the maximum cut problem,'' \emph{arXiv:1811.08419}, 2018.

\bibitem{goemans1995improved}
M.~X. Goemans and D.~P. Williamson, ``Improved approximation algorithms for
  maximum cut and satisfiability problems using semidefinite programming,''
  \emph{Journal of the ACM}, vol.~42, no.~6, pp. 1115--1145, 1995.

\bibitem{osti_1492737}
\BIBentryALTinterwordspacing
O.~D. Parekh, C.~Ryan-Anderson, and S.~Gharibian, ``Quantum optimization and
  approximation algorithms.'' Tech. Rep., 2019. [Online]. Available:
  \url{https://doi.org/10.2172%2F1492737}
\BIBentrySTDinterwordspacing

\bibitem{PhysRevA.97.022304}
Z.~Wang, S.~Hadfield, Z.~Jiang, and E.~G. Rieffel, ``Quantum approximate
  optimization algorithm for maxcut: A fermionic view,'' \emph{Physical Review
  A}, vol.~97, p. 022304, 2018.

\bibitem{Guerreschi2019}
\BIBentryALTinterwordspacing
G.~G. Guerreschi and A.~Y. Matsuura, ``{QAOA} for max-cut requires hundreds of
  qubits for quantum speed-up,'' \emph{Scientific Reports}, vol.~9, no.~1, May
  2019. [Online]. Available: \url{https://doi.org/10.1038/s41598-019-43176-9}
\BIBentrySTDinterwordspacing

\bibitem{2006PNAS..103.8577N}
M.~E.~J. {Newman}, ``From the cover: {M}odularity and community structure in
  networks,'' \emph{Proceedings of the National Academy of Science}, vol. 103,
  pp. 8577--8582, 2006.

\bibitem{brandes2006maximizing}
U.~Brandes, D.~Delling, M.~Gaertler, R.~G{\"o}rke, M.~Hoefer \emph{et~al.},
  ``Maximizing modularity is hard,'' \emph{arXiv:physics/0608255}, 2006.

\bibitem{palla2005uncovering}
G.~Palla, I.~Der{\'e}nyi, I.~Farkas, and T.~Vicsek, ``Uncovering the
  overlapping community structure of complex networks in nature and society,''
  \emph{Nature}, vol. 435, no. 7043, p. 814, 2005.

\bibitem{su2010glay}
G.~Su, A.~Kuchinsky, J.~H. Morris, D.~J. States, and F.~Meng, ``{GL}ay:
  community structure analysis of biological networks,'' \emph{Bioinformatics},
  vol.~26, no.~24, pp. 3135--3137, 2010.

\bibitem{bardella2016hierarchical}
G.~Bardella, A.~Bifone, A.~Gabrielli, A.~Gozzi, and T.~Squartini,
  ``Hierarchical organization of functional connectivity in the mouse brain:
  {A} complex network approach,'' \emph{Scientific Reports}, vol.~6, p. 32060,
  2016.

\bibitem{nicolini2017community}
C.~Nicolini, C.~Bordier, and A.~Bifone, ``Community detection in weighted brain
  connectivity networks beyond the resolution limit,'' \emph{Neuroimage}, vol.
  146, pp. 28--39, 2017.

\bibitem{negre2018}
C.~F. Negre, H.~Ushijima-Mwesigwa, and S.~M. Mniszewski, ``Detecting multiple
  communities using quantum annealing on the {D-Wave} system,''
  \emph{arXiv:1901.09756}, 2019.

\bibitem{kandala2017hardware}
A.~Kandala, A.~Mezzacapo, K.~Temme, M.~Takita, M.~Brink \emph{et~al.},
  ``Hardware-efficient variational quantum eigensolver for small molecules and
  quantum magnets,'' \emph{Nature}, vol. 549, no. 7671, p. 242, 2017.

\bibitem{farhi2017quantum}
E.~Farhi, J.~Goldkanstone, S.~Gutmann, and H.~Neven, ``Quantum algorithms for
  fixed qubit architectures,'' \emph{arXiv:1703.06199}, 2017.

\bibitem{guerreschi2017practical}
G.~G. Guerreschi and M.~Smelyanskiy, ``Practical optimization for hybrid
  quantum-classical algorithms,'' \emph{arXiv:1701.01450}, 2017.

\bibitem{bergholm2018pennylane}
V.~Bergholm, J.~Izaac, M.~Schuld, C.~Gogolin, and N.~Killoran, ``Pennylane:
  Automatic differentiation of hybrid quantum-classical computations,''
  \emph{arXiv:1811.04968}, 2018.

\bibitem{romero2018strategies}
J.~Romero, R.~Babbush, J.~McClean, C.~Hempel, P.~Love, and A.~Aspuru-Guzik,
  ``Strategies for quantum computing molecular energies using the unitary
  coupled cluster ansatz,'' \emph{Quantum Science and Technology}, 2018.

\bibitem{harrow2019low}
A.~Harrow and J.~Napp, ``Low-depth gradient measurements can improve
  convergence in variational hybrid quantum-classical algorithms,''
  \emph{arXiv:1901.05374}, 2019.

\bibitem{nelder1965simplex}
J.~A. Nelder and R.~Mead, ``A simplex method for function minimization,''
  \emph{The Computer Journal}, vol.~7, no.~4, pp. 308--313, 1965.

\bibitem{verdon2017quantum}
G.~Verdon, M.~Broughton, and J.~Biamonte, ``A quantum algorithm to train neural
  networks using low-depth circuits,'' \emph{arXiv:1712.05304}, 2017.

\bibitem{otterbach2017unsupervised}
J.~Otterbach, R.~Manenti, N.~Alidoust, A.~Bestwick, M.~Block \emph{et~al.},
  ``Unsupervised machine learning on a hybrid quantum computer,''
  \emph{arXiv:1712.05771}, 2017.

\bibitem{wecker2016training}
D.~Wecker, M.~B. Hastings, and M.~Troyer, ``Training a quantum optimizer,''
  \emph{Physical Review A}, vol.~94, no.~2, p. 022309, 2016.

\bibitem{yang2017optimizing}
Z.-C. Yang, A.~Rahmani, A.~Shabani, H.~Neven, and C.~Chamon, ``Optimizing
  variational quantum algorithms using {P}ontryagin's minimum principle,''
  \emph{Physical Review X}, vol.~7, no.~2, p. 021027, 2017.

\bibitem{zhu2018training}
D.~Zhu, N.~Linke, M.~Benedetti, K.~Landsman, N.~Nguyen \emph{et~al.},
  ``Training of quantum circuits on a hybrid quantum computer,''
  \emph{arXiv:1812.08862}, 2018.

\bibitem{brandao2018fixed}
F.~G. Brandao, M.~Broughton, E.~Farhi, S.~Gutmann, and H.~Neven, ``For fixed
  control parameters the quantum approximate optimization algorithm's objective
  function value concentrates for typical instances,'' \emph{arXiv:1812.04170},
  2018.

\bibitem{matsumine2019channel}
T.~Matsumine, T.~Koike-Akino, and Y.~Wang, ``Channel decoding with quantum
  approximate optimization algorithm,'' \emph{arXiv:1903.02537}, 2019.

\bibitem{nakanishi2019sequential}
K.~M. Nakanishi, K.~Fujii, and S.~Todo, ``Sequential minimal optimization for
  quantum-classical hybrid algorithms,'' \emph{arXiv:1903.12166}, 2019.

\bibitem{Grove}
\BIBentryALTinterwordspacing
Rigetti, ``Grove,'' 2019. [Online]. Available:
  \url{https://github.com/rigetti/grove}
\BIBentrySTDinterwordspacing

\bibitem{mcclean2016theory}
J.~R. McClean, J.~Romero, R.~Babbush, and A.~Aspuru-Guzik, ``The theory of
  variational hybrid quantum-classical algorithms,'' \emph{New Journal of
  Physics}, vol.~18, no.~2, p. 023023, 2016.

\bibitem{Powell2009a}
\BIBentryALTinterwordspacing
M.~J.~D. Powell, ``The {BOBYQA} algorithm for bound constrained optimization
  without derivatives,'' University of Cambridge, Tech. Rep. DAMTP 2009/NA06,
  2009. [Online]. Available:
  \url{http://www.damtp.cam.ac.uk/user/na/NA_papers/NA2009_06.pdf}
\BIBentrySTDinterwordspacing

\bibitem{RinnooyKan1987}
A.~H.~G. {Rinnooy Kan} and G.~T. Timmer, ``Stochastic global optimization
  methods, part {I: Clustering} methods,'' \emph{Mathematical Programming},
  vol.~39, no.~1, pp. 27--56, 1987.

\bibitem{RinnooyKan1987a}
------, ``Stochastic global optimization methods, part {II: Multi} level
  methods,'' \emph{Mathematical Programming}, vol.~39, no.~1, pp. 57--78, 1987.

\bibitem{Qiskit}
G.~Aleksandrowicz, T.~Alexander, P.~Barkoutsos, L.~Bello, Y.~Ben-Haim
  \emph{et~al.}, ``Qiskit: An open-source framework for quantum computing,''
  2019.

\bibitem{powell2009bobyqa}
M.~J. Powell, ``The {BOBYQA} algorithm for bound constrained optimization
  without derivatives,'' \emph{Cambridge NA Report NA2009/06, University of
  Cambridge, Cambridge}, pp. 26--46, 2009.

\bibitem{powell2006newuoa}
M.~J.~D. Powell, ``The {NEWUOA} software for unconstrained optimization without
  derivatives,'' in \emph{Large-scale Nonlinear Optimization}.\hskip 1em plus
  0.5em minus 0.4em\relax Springer, 2006, pp. 255--297.

\bibitem{brent2013algorithms}
R.~P. Brent, \emph{Algorithms for minimization without derivatives}.\hskip 1em
  plus 0.5em minus 0.4em\relax Courier Corporation, 2013.

\bibitem{rowan1991functional}
T.~H. Rowan, ``Functional stability analysis of numerical algorithms.'' Ph.D.
  dissertation, {U}niversity of {T}exas at {A}ustin, 1990.

\bibitem{libEnsemble_0.5.0}
\BIBentryALTinterwordspacing
S.~Hudson, J.~Larson, S.~M. Wild, and D.~Bindel, ``{libEnsemble} users
  manual,'' 2019. [Online]. Available:
  \url{https://buildmedia.readthedocs.org/media/pdf/libensemble/latest/libensemble.pdf}
\BIBentrySTDinterwordspacing

\bibitem{fortunato2010community}
S.~Fortunato, ``Community detection in graphs,'' \emph{Physics Reports}, vol.
  486, no. 3-5, pp. 75--174, 2010.

\bibitem{code}
\BIBentryALTinterwordspacing
 [Online]. Available:
  \url{https://github.com/rsln-s/Multistart-Methods-for-Quantum-Approximate-Optimization}
\BIBentrySTDinterwordspacing

\bibitem{chung1997spectral}
F.~R. Chung and F.~C. Graham, \emph{Spectral graph theory}.\hskip 1em plus
  0.5em minus 0.4em\relax American Mathematical Soc., 1997, no.~92.

\bibitem{nadakuditi2012graph}
R.~R. Nadakuditi and M.~E. Newman, ``Graph spectra and the detectability of
  community structure in networks,'' \emph{Physical Review Letters}, vol. 108,
  no.~18, p. 188701, 2012.

\bibitem{anschuetz2019variational}
E.~Anschuetz, J.~Olson, A.~Aspuru-Guzik, and Y.~Cao, ``Variational quantum
  factoring,'' in \emph{International Workshop on Quantum Technology and
  Optimization Problems}.\hskip 1em plus 0.5em minus 0.4em\relax Springer,
  2019, pp. 74--85.

\bibitem{Babbush2018}
\BIBentryALTinterwordspacing
R.~Babbush, N.~Wiebe, J.~McClean, J.~McClain, H.~Neven, and G.~K.-L. Chan,
  ``Low-depth quantum simulation of materials,'' \emph{Physical Review X},
  vol.~8, no.~1, Mar. 2018. [Online]. Available:
  \url{https://doi.org/10.1103/physrevx.8.011044}
\BIBentrySTDinterwordspacing

\end{thebibliography}

% \section*{Government License}
% The submitted manuscript has been created by UChicago Argonne, LLC, Operator of
% Argonne National Laboratory (``Argonne''). Argonne, a U.S. Department of Energy
% Office of Science laboratory, is operated under Contract No. DE-AC02-06CH11357.
% The U.S. Government retains for itself, and others acting on its behalf, a
% paid-up nonexclusive, irrevocable worldwide license in said article to
% reproduce, prepare derivative works, distribute copies to the public, and
% perform publicly and display publicly, by or on behalf of the Government.  The
% Department of Energy will provide public access to these results of federally
% sponsored research in accordance with the DOE Public Access Plan.
% http://energy.gov/downloads/doe-public-access-plan.

\end{document}